\documentstyle[epsfig,epsf]{article}

\def\Journal#1#2#3#4{{#1} {\bf #2}, #3 (#4)}

\def\PLB{{\em Phys. Lett.}  B}
\def\PRL{\em Phys. Rev. Lett.}
\def\PRD{{\em Phys. Rev.} D}

\begin{document}

\begin{center}
{\Large {\bf RESULTS FROM HIGH ENERGY ACCELERATORS}}
\end{center}

\vskip .7 cm

\begin{center}
G. GIACOMELLI and B. POLI \par~\par
{\it Dept of Physics, Univ. of Bologna and INFN, \\
V.le C. Berti Pichat 6/2, Bologna, I-40127, Italy\\} 

E-mail: giacomelli@bo.infn.it , barbara.poli@bo.infn.it

\par~\par

Lectures at the 6$^{th}$ School on Non-Accelerator Astroparticle Physics,  
\\ Trieste, Italy, 9-20 July 2001. 

\vskip .7 cm
{\large \bf Abstract}\par
\end{center}

{\normalsize We review some of the recent experimental results obtained at 
high-energy colliders with emphasis on LEP and SLC results.}

\vspace{5mm}

\section{Introduction}

High energy colliders allow to study two particle 
collisions at the highest energies, since the c.m. energy $\sqrt{s}$
grows linearly with the beam energy E$_{b}$: $\sqrt{s} = 2 E_{b}$.
In particular $e ^{+ }e^{- }$ collisions offer the possibility of studying in the 
simplest way the fundamental particles and their basic interactions\cite{results_from_acc_exp}.
The highest energy collider is the Tevatron at Fermilab with c.m. energies up to 
$2$ TeV. The highest energy $e^{+ }e^{- }$ collider was 
LEP2 at CERN; it was closed in november 2000. LEP, the SLC linear collider 
at SLAC and the Tevatron tested the Standard Model (SM) of 
electroweak and strong 
interactions to unprecedented precisions 
\cite{results_from_acc_exp}-\cite{budapest}.

Besides energy, the second important parameter of a collider is
its luminosity $\cal L$, which is defined as that number which multiplied 
by a cross-section $\sigma$ gives the collision rate $N:~~ N = \cal L\sigma$.
The highest energy $e^{+ }e^{- }$ colliders have or had luminosities in the 
range $10^{31}<\cal L$ $<10^{32}$  $ cm^{- 2} s^{- 1}$, which yield collision 
rates of $ \sim $ 1 event/s at  $\sqrt{s} = \mathrm{m}_Z$ and 
$ \sim $0.01 event/s at LEP2 ($130 <\sqrt{s}< 208$ GeV). Recent 
e$^{+ }$e$^{- }$ b-factories have much larger luminosities.

In these notes we shall first summarize the legacy 
of LEP to high energy physics and shall then discuss
some results from other high energy accelerators.

\begin{figure}[tbh]
\begin{center}
 \mbox{
   \epsfig{file=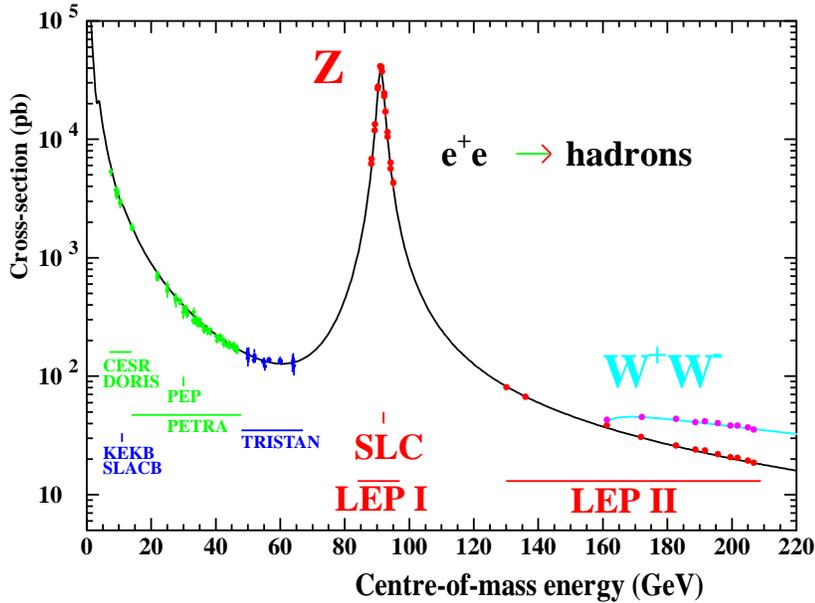,height=8.truecm,
            bbllx=140,bblly=95,bburx=490,bbury=330}
 }
\end{center}
\caption{Energy dependence of
$\sigma (e^+e^-\rightarrow$ hadrons).}
\label{fig:en_dep_sigma}
\end{figure}

Fig. \ref{fig:en_dep_sigma} shows the cross--section for 
$e^+e^-\rightarrow$ hadrons vs $\sqrt{s} $. 
Precise measurements 
of the $Z^0$ mass, width, and decay properties yield precision tests of the 
electroweak (EW) theory. In the SM the $ Z^{0}$ decay width is related 
to the number of fermion pairs into which it can decay. 
The more ways in which it can decay, the faster it decays and the wider the
$Z^0$ peak becomes.
Measurements of the $Z^0$ width constrain the number of generations
(3 neutrino families),
and deviations from an integer value may hint at new physics.
Radiative corrections modify the Breit--Wigner shape of the $Z^0$ resonance.
The analyses of radiative corrections yielded below threshold a precise value for the
top--quark mass,
in excellent agreement with the directly measured value \cite{LEP}.
\par
Since the $Z^0$ decays predominantly into quark--antiquark pairs,
it yields a clean
data sample with which to test quantum chromodynamics (QCD), the theory of the
strong interaction.
The $q\bar q$ pair is never observed directly, but
it gives rise to two opposite jets of hadrons by a process called
fragmentation (hadronization).
Before fragmentation takes place, one of the quarks may radiate
a gluon by a process similar to bremsstrahlung.
In this case, three jets of
hadrons are produced.
The ratio of the number of three--jet events to the number
of two--jet events is one way of measuring $\alpha_s$, the strong
coupling constant, which is a fundamental parameter of QCD.
\par
In addition to the quarks, the leptons and the vector bosons,
the SM requires at least one scalar Higgs particle
(the $H^0_{ SM}$ ), which is
needed for the hypothesized mechanism for the generation of masses.
The
coupling of the Higgs particle is predicted by the theory, but not its mass.
The precision measurements at the $Z^0$ yield indirect 
information on the $H^0_{ SM}$ mass.
A direct limit and a hint of a Higgs boson were 
obtained at LEP2.
\par
The higher energies and luminosities at LEP2 allowed to study
the triple boson vertex in 
$e^+e^-\rightarrow Z^0\rightarrow W^+W^-$,
measure with precision the mass and the width of $W^\pm$,
perform searches for new particles.
\par
Experiments at the  
HERA $e ^{+ }p$ collider provided information on the proton structure 
at the lowest x-values and largest Q-values;
they studied the photon structure, heavy flavour production, 
leading proton physics,
hadronic final states and performed many searches for new particles.
\par
Eperiments at the Fermilab $p \bar p $ collider found the top quark 
above threshold, provided precise measurements of $m_t$ and $m_W$,
studied  
quark and gluon collisions at the highest energies, yielding large $p_t$
events, studied minimum bias, low $p_t$ physics and performed many 
new particle searches.
\par
Recent experiments at fixed target accelerators gave information
on direct CP violation in the $K^{0} \bar{K}^{0}$ system \cite{CP-K},
while experiments at b-factories gave the first indications of CP 
violation in the $B^{0} \bar{B}^{0}$ system \cite{CP-B}.
\par
Among the very many results obtained in fixed target experiments 
at accelerators, we mention very briefly: the 
study of Deep Inelastic Scattering (DIS) with longitudinally 
polarized $\mu^-$ and $e^-$ on protons
in sophisticated experiments at CERN and SLAC; 
they found that only $30\%$ of the proton spin is carried 
by the valence quarks; it would seem that the main fraction of the
proton spin is carried by gluons \cite{spin}.

\section{Experimental}
Most of the recent high energy physics data come from high energy colliders,
where several 
large, 4$\pi$, general purpose detectors
may operate simultaneously.
Each of them is made of many
subdetectors, whose combined role is to  measure the energy, 
direction, charge, and type of the particles produced.
Apart from neutrinos, no particle should be able to escape a detector
without leaving some sign of its passage.
Each subdetector has a cylindrical structure, with a ``barrel" and two
``end-caps".
Tracking is performed by a central
detector, whilst electron and photon energy measurements are carried out by a
high--resolution electromagnetic calorimeter;
the magnet iron yoke is instrumented as a hadron calorimeter;
it is followed by a muon detector.
At $e^+e^-$ colliders, a forward detector 
completes the e.m. coverage by tagging
small--angle electrons and photons, and is used as a precision
luminosity monitor. 
\par
In order to select and measure the cross--section for a specific channel,
one needs: (i) a trigger, (ii) the required events $(N_i)$,
(iii) the computation of the global efficiency $(\epsilon_i)$, and
(iv) a luminosity determination $(\int \cal{ L}$$~ dt$):
$\sigma_i = N_i/(\epsilon_i~\int \cal {L}$$~ dt)~$.
Recent improvements concern the vertex detectors, in view of 
studying short lived particles and for direct searches for the 
$H^0_{ SM}$, for example in 
$e^+e^-\rightarrow Z^0\rightarrow Z^0+H^0_{ SM}  \rightarrow Z^0 + b\bar{b}$.

\section{$e^+e^-$ collisions. The LEP legacy}
\subsection{Electroweak physics}
At energies around the $Z^0$ peak the basic processes are
\begin{equation}
{\mathrm e}^+{\mathrm e}^- \rightarrow Z^0,\gamma
\rightarrow f\bar f~,~~~~~~~
f\bar f=q\bar q,\ell\bar \ell,
\end{equation}
The $\ell\bar \ell$ pairs may be charged
(e$^+$e$^-$, $\mu^+\mu^-,\, \tau^+\tau^-$) or neutral
($\nu_{\mathrm e}\bar \nu_{\mathrm e}$,
$\nu_\mu\bar \nu_\mu$, $\nu_\tau\bar \nu_\tau$) leptons.
The $q\bar q$ pairs are u\=u, d\=d, s\=s, c\=c, and b\=b (the
t\=t has a higher energy threshold).
Each $q$ or $\bar q$ hadronizes in a jet of hadrons.
Thus the $q\bar q$ pairs are characterized by two opposite jets of
hadrons.
The $q$ (or the $\bar q$) may radiate a gluon, which yields a third jet.
\par
The behaviour of any cross--section around the $Z^0$ peak is
typical of a resonant state
with $J$ = 1, described by a relativistic Breit--Wigner
formula, plus an electromagnetic term and an interference term: 
\begin{equation}
\begin{array}{lccccc}
\sigma (e^+e^-\rightarrow f\bar f)= &
{4\over 3}\pi{{\alpha (m^2_Z)}\over s} & + & I{{s-m^2_Z}\over s} & + &
{{12\pi}\over {m^2_Z}}\Gamma _{e^+e^-}\Gamma _{f\bar f}
{s\over {(s-m^2_Z)^2+{{s^2}\over {m^2_Z}}\Gamma^2_Z}} \\
& \mbox{electrom.}    && \mbox{interf.}   && \mbox{resonant} \\
& \mbox{term}       && \mbox{term}      && \mbox{term}
\end{array}
\label{eq:Breit-Wigner}
\end{equation}
This formula has to be convoluted with initial state radiation.
Around the $Z^0$, the first two terms of Eq. \ref{eq:Breit-Wigner}
are small corrections to the main term,
which is the $Z^0$ Breit--Wigner.
$\Gamma_{f\bar f}$ is the partial width for the decay of the
$Z^0$ into a fermion--antifermion pair, $Z^0\rightarrow f\bar f$.
The total width $\Gamma_Z$ is given by
\begin{equation}
\Gamma_Z = \Gamma_{\mathrm h} + \Gamma_{\mathrm {e}} + \Gamma_{\mu} +
\Gamma_{\tau} + N_\nu\Gamma_{\nu} = \Gamma _{\mathrm {vis}} + \Gamma _{\mathrm
{inv}}~,
\end{equation}
where $\Gamma_{\mathrm h}$ is the hadronic width and $\Gamma_{\mathrm
{e}}, \Gamma_{\mu}, \Gamma_{\tau}, \Gamma_{\nu}$ are the
leptonic widths; it is costumary to use
$R_e =\Gamma_{\mathrm{h}} /\Gamma_{\mathrm{e}}$,
$R_\mu =\Gamma_{\mathrm{h}} /\Gamma_\mu$ and
$R_\tau =\Gamma_{\mathrm{h}} /\Gamma_\tau$.
In the Standard Model
\begin{equation}
\left\{
\begin{array}{lcl}
\Gamma_{\mathrm h} &=& \Gamma_{\mathrm u} + \Gamma_{\mathrm d} +
    \Gamma_{\mathrm s} + \Gamma_{\mathrm c} + \Gamma_{\mathrm b} =
    1742~{\mathrm{MeV}}\\
\Gamma_{\mathrm u} &=& \Gamma_{\mathrm c} = 296~{\mathrm{MeV}}~~,
    ~~~~ \Gamma_{\mathrm d} =
    \Gamma_{\mathrm s} = \Gamma_{\mathrm b} = 374~{\mathrm{MeV}}\\
\Gamma_{\mathrm {e}} &=&
    \Gamma_{\mu} = \Gamma_{\tau} = 84.0~{\mathrm{MeV}}\\
\Gamma_{\nu}&=& 167.3\ {\mathrm{MeV}}~~,
    ~~~~ N_\nu = 3~~,
    ~~~~ \Gamma_{\mathrm{inv}} =501.6~{\mathrm{MeV}}\\
\Gamma_{\mathrm Z} &=& \Gamma_{\mathrm h} + 3\Gamma_{\ell} +
3\Gamma_{\nu}= 2495~{\mathrm{MeV}}~.
\end{array}
\right.
\end{equation}
\par
In the electroweak theory each partial width $\Gamma _f$
may be expressed
in terms of vector ($g_v$) and axial--vector ($g_a$) coupling constants
\begin{equation}
g_a = I_{3{\mathrm f}}~~~,~~~~g_v =
I_{3{\mathrm f}} - 2 Q_{\mathrm f} \sin^2 \theta_{\mathrm w}~,
\end{equation}
where $I_{3{\mathrm f}}$ is the third component of the weak isospin,
$Q_{\mathrm f}$ the  electric charge of the fermion and $\theta_{\mathrm w}
$
the weak mixing angle:
\begin{equation}
\Gamma _f = N_{\mathrm c}{G_\mu~m^3_Z\over 6\pi
\sqrt{2}}(g_v^2+g_a^2)(1+\delta _f)
\label{eq:Gamma}
\end{equation}
$N_{\mathrm c}$ is the number of colours $(N_{\mathrm c}$ = 1 for
leptons, $N_{\mathrm c}$ = 3 for quarks), $\delta _f$
 accounts for QED+EW corrections.
Angular distributions of the produced $f\bar f$ exhibit
asymmetries; at $m_Z$ they are written as
\begin{equation}
A_{\mathrm {FB}}^{0,f}=\frac{3}{4}
A_{\mathrm {e}} A_f ~~~,~~~~~
A_f=\frac{2g_v g_a}{g_v^2 + g_a^2}~.
\end{equation}
\begin{figure}[thb]
\begin{center}
 \mbox{
    \epsfig{file=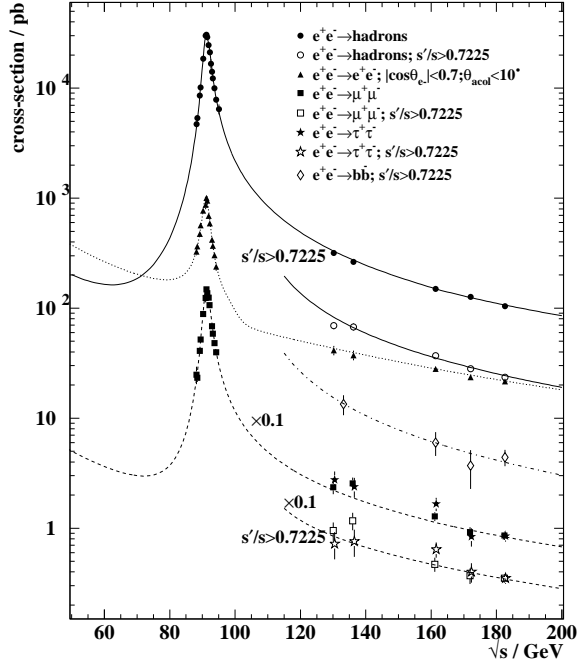,height=9.truecm,
            bbllx=0,bblly=80,bburx=570,bbury=690}
 }
\end{center}
\caption{e$^+$e$^-$ cross sections vs c.m. energy.}
\label{fig:crosssection-vs-cm}
\end{figure}
Fig. \ref{fig:crosssection-vs-cm}
shows the energy dependence of $\sigma (e^+e^-\rightarrow$
hadrons$(\gamma))$
and of $\sigma(e^+e^-\rightarrow l^+l^-(\gamma))$,
without removing, and removing, radiative events;
the SM predictions (lines) are in agreement with the data (points)
\cite{PN}.
\par
At energies $ \sqrt{s} > 100$ GeV the probability of radiating 
a photon by the initial $e^+$ or $e^-$ is high; thus the effective 
collision energy $ \sqrt{s'} < \sqrt{s}$; the distribution in $ \sqrt{s'}$ 
indicates clearly a peak at the $Z^0$ resonance. 

\subsection{Radiative corrections}
\begin{figure}[thb]
\begin{center}
 \mbox{
    \epsfig{file=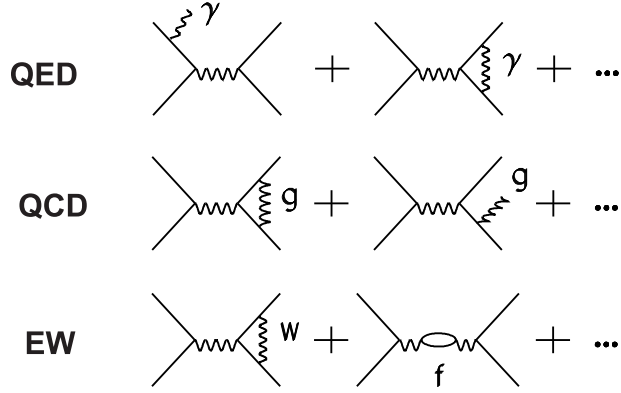,height=6.truecm,
             bbllx=65,bblly=355,bburx=520,bbury=655}                    
 }
\end{center}
\caption{Radiative correction diagrams.}
\label{fig:marcellini}
\end{figure}

Radiative corrections for the processes
 $e^+e^-\rightarrow f\bar f $
include QED, EW and QCD corrections, which may be summarized 
by the Feynman diagrams of Fig. \ref{fig:marcellini}. 
\\
{\bf QED radiative corrections } : (i) photon emission from the initial state 
$e^+$, $e^- $ modifies the effective c.m. 
energy and distorts the Breit--Wigner shape of the $Z^0$ resonance.
(ii) QED radiative corrections + EW and QCD corrections lead to the 
running of the electromagnetic coupling costant  
\begin{equation}
\alpha _{\mathrm QED}(s) =  {\alpha _{\mathrm QED}(m^2_Z) \over  
{1 - \Delta \alpha _{\mathrm l}(s) - \Delta \alpha ^{(5)}_{\mathrm h} -
 \Delta \alpha _{\mathrm {top}}(s)}}
\end{equation}
where $\Delta \alpha _{\mathrm l}(s)$ and $\Delta \alpha _{\mathrm {top}}(s)$
are well known; $\Delta \alpha ^{(5)}_{\mathrm h}$ arises from the contribution 
of light quarks to the photon vacuum polarization;  
this correction may be computed using the new $e^+e^- $
data obtained by the BES experiment in China for energies below 12 GeV 
\cite{BES}.
The new determination
$\Delta \alpha ^{((5)}_{\mathrm h}(m^2_Z) = 0.02761\pm 0.00036$
has a smaller error than previous determinations \cite{burk}. On the other hand
the same new BES data plus more theoretical input in the low energy 
region lead to 
$\Delta \alpha ^{(5)}_{\mathrm h}(m^2_Z) = 0.02738\pm 0.00020$ 
\cite{martin}.\\
{\bf EW radiative corrections} : they lead to corrections to the coupling 
costants of the $Z^0$ to fermions. These corrections may be absorbed in the 
definitions (i) of the electroweak mixing angle 
\begin{equation}
\sin ^{2}\theta _{\mathrm W} = 1- \frac{m^{2}_{\mathrm W}}{m^{2}_{\mathrm Z}}
~~~\Rightarrow ~~~\sin ^{2} \theta^{\mathrm {lept}}_{\mathrm {eff}} = 
(1 + \Delta \mathrm K) \sin ^{2} \theta _{\mathrm W} 
\end{equation}
(ii) of the $ \rho $ parameter:
\begin{equation}
 \rho = \frac{m^{2}_{\mathrm W}}
{m^{2}_{\mathrm Z}\cos ^{2}\theta _{\mathrm W}} = 1 ~~~\Rightarrow ~~~
\tilde{\rho } = 1 + \Delta \rho 
\end{equation}
(iii) of $m_{\mathrm W}$:
\begin{equation}
m^{2} _{\mathrm W} = \frac{\pi \alpha }{\sqrt{2}\sin ^{2}\theta _{\mathrm W}
G_{\mathrm f}}(1 + \Delta r),  ~~~~~   \Delta r = 
\Delta \alpha _{\mathrm QED} + \Delta r_{\mathrm W}
\end{equation}
The corrections are therefore summarized in $\Delta K$, $\Delta \rho$, 
$\Delta  r_{\mathrm W}$, which depend quadratically on $m_{\mathrm {top}}$
and logarithmically on  $m_{\mathrm {H^{0}}}$.\\
{\bf QCD radiative corrections} : they are difficult to 
compute; different techniques must be used for each 
particular problem considered. For example, perturbation 
theory of a fixed order is adequate for the total cross section.
Neglecting strong corrections and quark masses,
the total hadronic width of the $Z^0$ boson can be written as 
$\Gamma_{\mathrm h} =  \Sigma _q ~\Gamma_q$
where $\Gamma_q$ is given in Eq.  \ref{eq:Gamma}.
The strong radiative corrections, to order $\alpha^3_s$,
modify  $\Gamma_{\mathrm h}$, which can now be expressed as
\begin{equation}
\Gamma'_{\mathrm h} = \Gamma_{\mathrm h} \left[1+c_1\frac{\alpha_s}{\pi}
+c_2(\frac{\alpha_s}{\pi})^2
+c_3(\frac{\alpha_s}{\pi})^3\right],
\end{equation}
where $c_1$, $c_2$ and $c_3$ have been calculated.
\par
Fixed order perturbation theory is not adequate for QCD 
calculations in the back--to--back region
where large logarithmic factors can arise, destroying the convergence
of the perturbative expansion.
Resummation at all orders of perturbation theory is necessary and 
was done for different variables.

\subsection{Precision measurements}
At each energy around the Z$^0$ peak,
measurements of the cross--sections have been carried out for
$e^+e^- \to Z^0 \to$ hadrons,
$e^+e^-$,
$\mu^+\mu^-$, $\tau^+\tau^-$, the forward--backward lepton asymmetries
$A^{\mathrm{e}}_{\mathrm {FB}}$,
$A^\mu_{\mathrm {FB}}$, $A^\tau_{\mathrm {FB}}$, the $\tau$ polarization
asymmetry $P_\tau$, the b\=b and c\=c partial widths and
forward--backward asymmetries, and the $q\bar q$ charge asymmetries.
As time goes on more parameters are added.
The latest one is $m_W$, whose determination improves every year 
(see Table \ref{tab:W}).
These measurements plus informations
from $ \nu N$ and $ \overline{p}p $ collisions allow to make precise 
tests of the SM and constrain the Higgs boson mass.
In summary the measured quantities are:\\\\
\underline{From LEP1 + SLC }
\begin{center}
Mass and width of the $Z^{0} ~~~~~~~~~~~~~~~~~~~~~ m_{Z},~
\Gamma_Z = \Gamma_{\mathrm h} + \Gamma_{\mathrm {e}} + \Gamma_{\mu} +
\Gamma_{\tau} + N_\nu\Gamma_{\nu}$  
\end{center}
\medskip
$\left\{
\begin{array}{lcl}
\mathrm {Hadronic~ cross~ section ~at ~} \sqrt{s} = m_{Z^{0}}  
~~~~~    \sigma^{h} _{0}\\
\mathrm {Ratios~ at~}  \sqrt{s} = m_{Z^{0}}
   ~~~~~~~~~~~~~~~~~~~~~~~~~~  
R^{0}_{l} = \frac{\Gamma _{ h}}{\Gamma _{l}}~ (l = 
\mathrm e, \mu , \tau),~ R^{0}_{b},~ R^{0}_{c}\\
\mathrm {Asymmetries~ at~} \sqrt{s} = m_{Z^{0}} ~~~~~~~~~~~~~~~~~    
A^{0, l}_{FB} = 
(\frac{N_{F} - N_{B}}{N_{F} + N_{B}})^{0, l},  ~  A^{0, b\overline{b}}_{FB}, 
~  A^{0, c\overline{c}}_{FB}\\
\mathrm {The~ \tau~ polarization~ parameter} ~~~~~~~~~~~~~    P_{\tau }
\end{array} 
\right.
$
\medskip
\par\noindent
\underline{From SLC }
\begin{center}
The left-right asymmetry ~ A$^{0}_{LR}$ and the leptonic LR-FB 
asymmetry  ~ A$^{LR}_{FB}$\\
measured with polarized beams 
\end{center}
\medskip
\underline{From LEP2 + p$\bar p$}~~~~~~~~~~~~~~~~~~~~~~~~~~m$_{W}$\\\\
\underline{From p$\bar p$}  ~~~~~~~~~~~~~~~~~~~~~~~~~~~~~~~~~~~~~m$_{t}$\\

The LEP, SLC and $p\bar p$ experiments give to the LEP working 
group (WG) their measured values; the WG makes averages 
which are used in the fits.

\noindent
{\bf Early fits}: They yielded precise tests of 
\par\noindent
-- lepton universality:~~~~~ $R_e = R_\mu = R_\tau $, 
~~~ $ A_{e} = A_{\mu} = A_{\tau }$,
\par\noindent
-- number of neutrino families: ~~~~~$ N_{\nu } = 2.9841 \pm  0.0083$.
\par\noindent
{\bf Latest fits} \cite{LEP},\cite{budapest}: 
{\bf (i)} Fit without the measured values of m$_{W}$, m$_{t}$.
This fit ``below threshold" yields values of m$_{t}$, m$_{W}$, m$_{H^0_{SM}}$:
\begin{center}
$\mathrm m_{t} = 169.0  
\begin{array}{lcl}
+11.5\\
-8.9
\end{array} $
GeV,~~ $\mathrm m_{W} = 80.363 \pm 0.032$ GeV,~~
$\mathrm m_{H^0_{SM}} = 81 \begin{array}{lcl}
+109\\
-40
\end{array} $ GeV,
\end{center}
which can be compared with the measured values of m$_{t}$, m$_{W}$:
\begin{center}
$\mathrm m_{t} = 174.3 \pm 5.1$ GeV,~~  
$\mathrm m_{W} = 80.448 \pm 0.034$ GeV
\end{center}
The agreement of the direct and indirect determinations of m$_{t}$, m$_{W}$
shows the consistency of the SM.

\noindent
{\bf (ii)} Global fit \cite{LEP},\cite{budapest}:
 including all data (also  m$_{t}$, m$_{W}$)
and new $\Delta \alpha ^{(5)}_{\mathrm h}$ corrections. The fit has
$ \chi ^{2}/DoF = 22.9/15$ and yields the following values of 
m$_{t}$, m$_{W}$, m$_{H^0_{SM}}$:
\begin{center}
$\mathrm m_{t} = 175.8 \pm 4.3$ GeV,~~  
$\mathrm m_{W} = 80.398 \pm 0.019$ GeV,~~
$\mathrm m_{H^0_{SM}} = 88\begin{array}{lcl}
+53\\
-35
\end{array} $ GeV
\end{center}

\begin{figure}[thb]
\begin{center}
 \mbox{    
    \epsfig{file=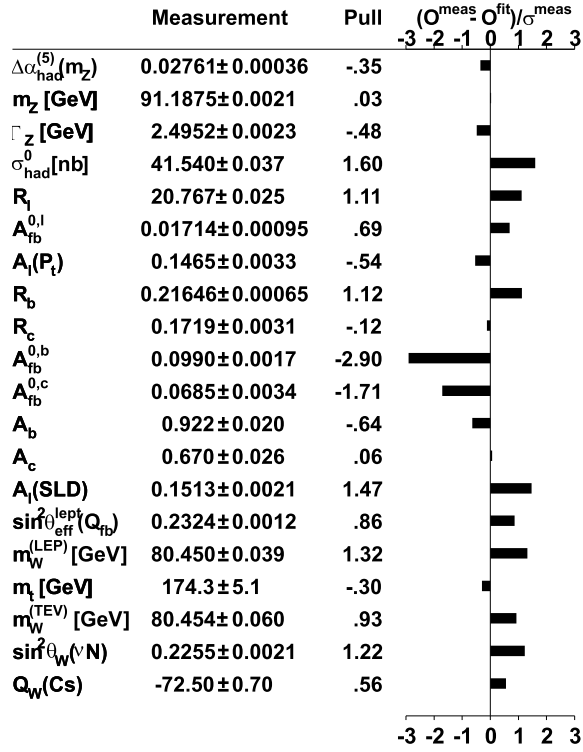,height=10.truecm,
            bbllx=0,bblly=70,bburx=594,bbury=790}
 }
\end{center}
\caption{Precise measurements of the electroweak 
parameters obtained from
the global fit, and the pull parametrs
= $ \left[O^{meas} - O^{fit} \right] /  \sigma ^{meas}$ \cite{budapest}.}
\label{fig:average}
\end{figure}

Fig. \ref{fig:average} gives the values of the quantities obtained from the global fit 
and the pull parameters. The fit is not very good; this comes mainly 
from A$^{0,b}_{FB}$: it could be a statistical fluctuation, an error in the
measurement of A$^{0,b}_{FB}$ or a hint of physics beyond the 
SM. Checks are been made on all the measurements of A$^{0,b}_{FB}$.
We point out that the present precision on 
$ \sin ^{2}\theta _{W}$, m$ _{W}$ and m$ _{Z}$ are almost two
orders of magnitude better than what anticipated in 1984. 
 
\begin{figure}[thb]
\begin{center}
 \mbox{
   \epsfig{file=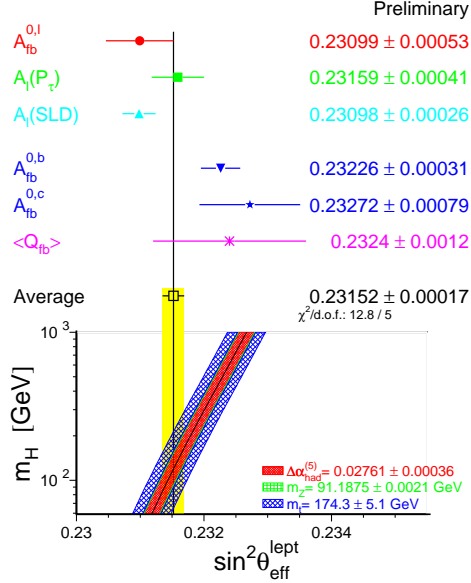,height=8.truecm,
            bbllx=0,bblly=0,bburx=594,bbury=730}
}
\end{center}
\caption{Determinations of 
$ \sin ^{2}\theta ^{lept}_{eff}$ from different asymmetries. 
Also shown is the SM 
prediction as function of m$_{H^0_{SM}}$.}
\label{fig:sin}
\end{figure}

Fig. \ref{fig:sin} shows a comparison of several determinations of 
$ \sin ^{2}\theta^{lept}_{eff}$ from measurements of 
different asymmetries; it also shows the prediction of the SM 
as a function of the Higgs boson mass.

\begin{figure}[thb]
\begin{center}
 \mbox{
   \epsfig{file=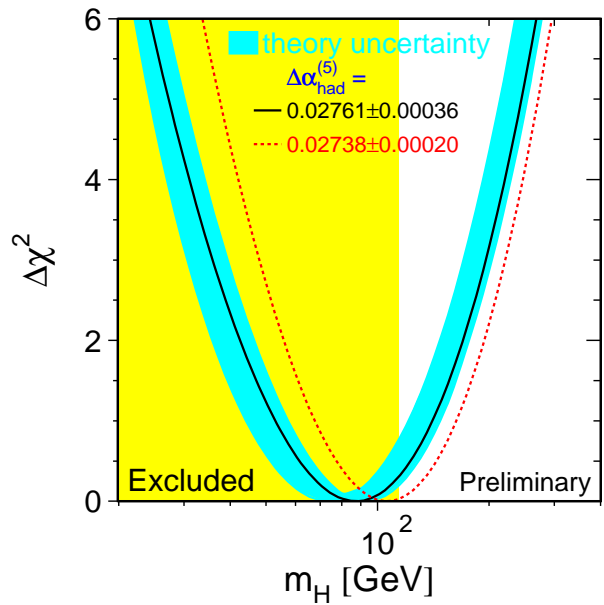,height=8.truecm,
             bbllx=0,bblly=35,bburx=594,bbury=565}
}
\end{center}
\caption{$\Delta \chi ^2 = \chi^2 - \chi^2_{min}$ vs m$_{H^0_{SM}}$. 
The solid line is the result of the fit using all data, the band rapresents
an estimate of the theoretical error due to missing higher order corrections.
The vertical band shows the 95\% CL exclusion limit on $m_{H}$ from
the direct search. The dashed curve is the result obtained using a different
evaluation of $\Delta \alpha ^{(5)}_{\mathrm h }(m^2_Z)$, see text.}
\label{fig:mh}
\end{figure}

\subsection{The SM Higgs boson }
Fig. \ref{fig:mh} shows the $ \chi ^{2}$ of the global fit as a function 
of $m_{H^0_{SM}}$ for two values of $\Delta \alpha ^{(5)}_{\mathrm h}$.
The fit indicates that 
at the 95\% CL the H$^{0}_{SM}$ mass should be lower than 196 GeV
\cite{LEP},\cite{budapest}.

The direct search for the SM Higgs boson is performed via the reaction 
$ e^{+}e^{-}\rightarrow Z^0 \rightarrow Z^0 + H^0_{SM}$, considering all 
possible decays for the $Z^0$ and $H^0_{SM}$ ($ Z^0 \rightarrow f\bar{f}$,
$ H^0_{SM} \rightarrow f'\bar{f'}$) \cite{particle-searches-at-lep}. 
The $H^0_{SM}$ prefers to decay into 
the heaviest particles; thus the channel $ H^0_{SM} \rightarrow b\bar{b}$
has the highest branching ratio. For this reason all microvertex detectors 
were upgraded for the LEP2 phase in order to cover the largest angular range 
for $b\bar{b}$ with the highest efficiency. The search suffers from SM
processes which simulate Higgs boson candidates. Complex procedures 
based on likelihood  methods have been deviced to cope with this background. 
The present situation for the $H^0_{SM}$ search at LEP2 may summarized 
saying that there is a signal for a preferred  mass of 115.6 GeV; the probability 
that the background may simulate the signal is 3.4\%.
Thus the indication (evidence) for a SM Higgs is at the level of about 
2.5 standard devations. From the same combination of data, a 
the 95\% CL lower bound is obtained: $m_{H^0_{SM}} > 114.1$ GeV 
\cite{higgs},\cite{Igo-Kemenes}.

\subsection{Multihadronic events }
Multihadron production in e$^+$e$^-$ annihilations
proceeds via four distinct phases.
\par\noindent
i) In the first phase the initial
e$^+$e$^-$ pair annihilates into a virtual $Z^0/\gamma$,
which yields a $q\bar q$ pair: this phase is 
described by the EW perturbative theory, and occurs at distances
of the order of 10$^{-17}$ cm.
Before annihilation, a $\gamma$ may be emitted by the initial e$^+$ or e$^-$,
thus reducing the effective c.m. energy.
\par\noindent
ii) In the second phase the $q$ (or the $\bar q$)
may radiate a gluon,
which subsequently radiates a second gluon
(yielding a three--gluon vertex), or a $q\bar q$ pair.
This phase, described by perturbative QCD,
occurs of distances $\sim 10^{-15}$ cm.
\par\noindent
iii) In the third phase the 
quarks and gluons fragment (ha\-dro\-ni\-ze) into  colourless hadrons
(this occurs over distances of $\sim 1$ fm).
This phase cannot be
analysed with perturbative methods;
it is treated with models.
\par\noindent
iv) In the 4th phase the produced hadron resonances decay
via strong
interaction (SI) into ``stable" hadrons (e.g. $\rho^0\to\pi^+\pi^-)$;
other hadrons decay via
the EM interaction $(\Sigma^0 \to \Lambda^0\gamma$,
$\pi^0 \to\gamma\gamma$);
b-hadrons decay via the weak
interaction (with lifetimes of about 10$^{-12}$ s).
This phase is described by
models which include experimental information on lifetimes
and branching ratios.
\begin{figure}[tb]
\begin{center}
 \mbox{
    \epsfig{file=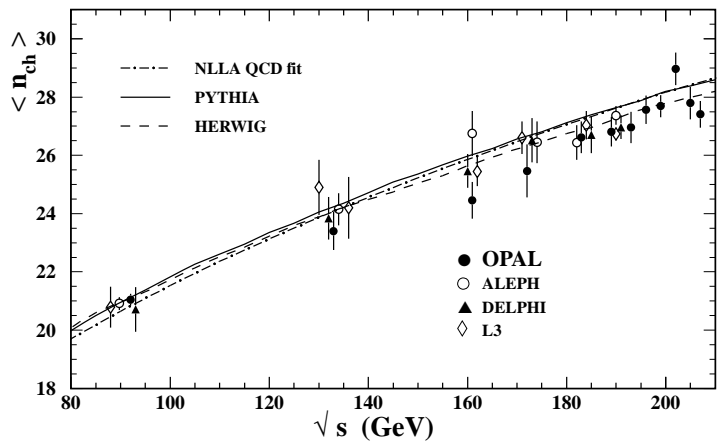,height=6truecm,
            bbllx=0,bblly=150,bburx=594,bbury=420}
 }
\end{center}
\caption{Average charged hadron multiplicity in e$^+$e$^-$ collisions vs.
c.m. energy.
The lines represent different predictions of the energy 
evolution of $\langle n_{ch}\rangle$.}
\label{fig:average-cha-mul}
\end{figure}
\par
Fig. \ref{fig:average-cha-mul}
shows the average charged hadron multiplicity, $\langle n_{ch}\rangle$,
  in e$^+$e$^-$ collisions plotted vs $ \sqrt{s}$.
At the $Z^0$ peak 
$\langle n_{ch}\rangle = 21.07 \pm 0.11$ \cite{PDG}, at 206 GeV  
$\langle n_{ch}\rangle \cong 27.7$ \cite{nmed}.
\par
A number of shape parameters (Sphericity, Thrust,...)
have been introduced to characterize the global event
structure of multihadronic final states. Their studies provide checks of QCD
and allow optimization of the Monte Carlos, which play
major roles for corrections and for analyses.
\par
For each charged track $k$ of a multihadronic  event, one defines the rapidity, the
transverse and longitudinal momenta and the variable
$x_k = p_k / E_{\mathrm {beam}}$.
\par
\begin{figure}[htb]
\begin{center}
 \mbox{
    \psfig{figure=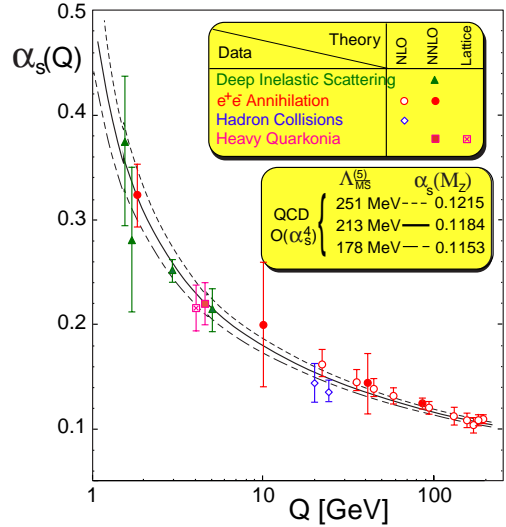,height=7truecm}
 }
\end{center}
\caption{$\alpha_s$ vs $Q=\sqrt{s}$~.
        The labels NLO and NNLO refer to the order of calculations used.
        NLO corresponds to O($\alpha_s^2$), and NNLO to O($\alpha_s^3$).
        The curves are O($\alpha_s^3$) QCD predictions.}
\label{fig:alpha}
\end{figure}

\noindent
{\protect\boldmath\bf {$\alpha_s$},
the coupling constant of the SI } is
a fundamental parameter
which may be determined from many types of measurements:
(i) from the ratio of 3-jet to 2-jet events,
(ii) from shape variables,
(iii) from
$~\Gamma_{\mathrm h}/\Gamma_\ell = R^0_Z (1 + \delta_{\mathrm{QCD}})~$,
and others.
LEP experiments established the flavour independence of
$\alpha_s$ and the decrease of $\alpha_s$ with increasing energy 
\cite{QCD1},\cite{QCD2},
Fig. \ref{fig:alpha}.
$\alpha_s (\mu)$ can be written as a function of
$\ln (\mu^2/\Lambda ^2)$, where $\Lambda$ is the QCD scale parameter
and $\mu$ is the renormalization scale
\begin{equation}
\alpha_s(\mu) = {{12\pi}\over {(33 - 2 n_f)\ \ln (\mu^2/\Lambda^2)}}
\left[ 1 - {{6(153 - 19 n_f)}\over {(33 - 2 n_f)^2}}\ {{\ln~[\ln~
(\mu^2/\Lambda^2)]}\over {\ln~ (\mu^2/\Lambda^2)}}\right]+...
\label{eq:alpha}
\end{equation}
$n_f$ is the number of active quarks with mass smaller 
than the energy scale $\mu$.
Eq. \ref{eq:alpha} predicts $\alpha_s \rightarrow 0$ as
$\mu \rightarrow \infty$ (asymptotic freedom).
With
increasing energy, $n_f$ changes
by discrete amounts as a new flavour threshold is crossed.
Also $\Lambda$ changes 
through a flavour threshold, $\Lambda \rightarrow \Lambda (n_f)$.
$\mu^2$ is usually chosen to be the c.m. energy ($\mu^2$ =
$E^2_{\mathrm{cm}}$);
a definition $\mu^2 = f \cdot E^2_{\mathrm{cm}}~,$
with 0.001 $< f <$ 0.01, gives a better description of the
jet production rates.

\par\noindent
{\bf Quark and gluon jet differences.}
QCD predicts different coupling strengths for the 
radiation of an additional gluon from either a quark or a gluon.
The coupling strengths are governed by the colour factors for gluon emission,
which have the values $C_F=4/3$ and $C_A=3$ for radiation from a quark and for a
gluon, respectively.
These are inclusive factors which need corrections to predict real jets.
QCD predicts that a gluon is more likely to radiate a gluon than a quark,
and that a gluon jet
has a higher particle multiplicity, a softer particle spectrum and is
broader than a quark jet of equal energy.
Many studies were done at LEP
to establish the differences between quark jets and gluon jets 
\cite{quark-gluon}.

\subsection{The reactions ${\protect\boldmath{e^+e^-\rightarrow W^+W^-}}$ and
${\protect\boldmath{e^+e^-\rightarrow Z^0Z^0}}$ }
\begin{figure}[htb]
\begin{center}
 \mbox{
   \epsfig{figure=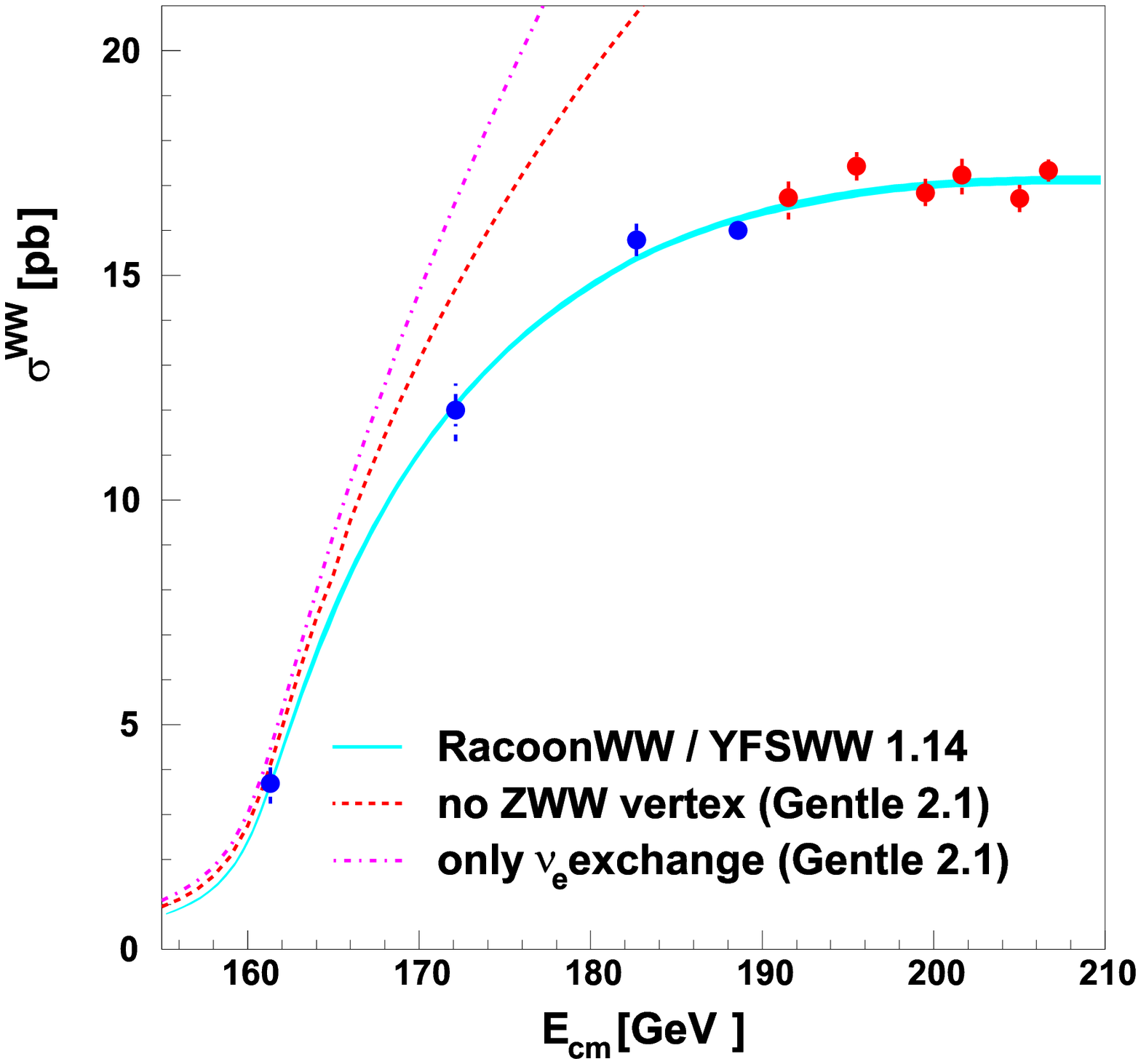,height=6.0truecm,
            bbllx=20,bblly=160,bburx=530,bbury=660}

   \epsfig{figure=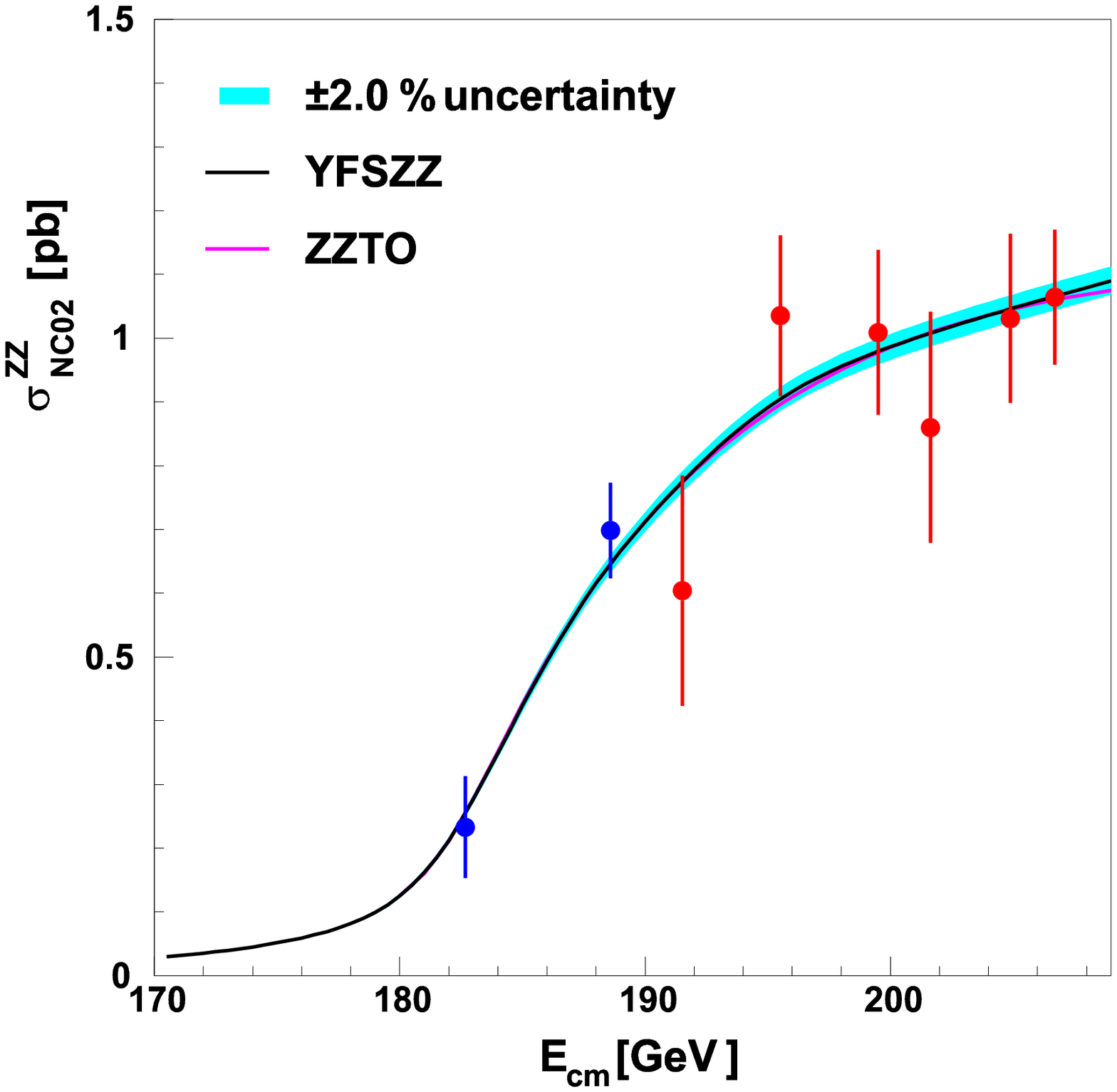,height=6.0truecm,
            bbllx=20,bblly=160,bburx=530,bbury=660}

 }
\end{center}
\caption{ (a) The total $e^+e^-\rightarrow W^+W^-$ 
        cross section vs $\sqrt{s}$ compared with the full 
        SM prediction and the predictions without the ZWW
        vertex and with only the t-channel $\nu_e$ exchange 
        diagram. (b) Cross sections for the reaction
        $e^+e^-\rightarrow Z^0Z^0$ vs $\sqrt{s}$.}
\label{fig:eeWWZZ}
\end{figure}

For $\sqrt{s}>2m_W$ one can study 
the $~e^+e^-\rightarrow W^+W^-~$reaction,
perform precision $W$-physics and test the SM triple
boson vertex $ZWW$.
The most important diagrams for
$W$-pair production are the s-channel $\gamma$/Z exchange, and the t-channel
neutrino exchange.
Fig. \ref{fig:eeWWZZ}a shows the energy dependence of the
$e^+e^-\rightarrow W^+W^-$ cross section 
\cite{budapest},\cite{WW}:
it has the tipical dependence of a
reaction just above threshold. The measured values are 
consistent with the SM predictions.
In a specific model there are 3 independent trilinear
gauge couplings
(anomalous couplings), which could affect both the total production
cross section and the shape of the differential cross-section. 
Since no deviations from SM predictions have 
been observed, one can 
place only upper limits for anomalous couplings.
\par
The results of direct measurements of $m_W$ are given in
Table \ref{tab:W}.
\begin{table}[tbh]
\caption{Summary of direct $W$ mass measurements.}
\label{tab:W}
\small
\begin{center}
\begin{tabular}{||c|c||}
\hline
p$\bar p$ colliders            &$80.454 \pm 0.060$ GeV\\
LEP2                             &$80.450 \pm 0.039$\\
\hline
Average                         &$80.451 \pm 0.033$\\
\hline
\end{tabular}
\end{center}
\end{table}

Fig. \ref{fig:eeWWZZ}b shows the energy dependence of
$\sigma (e^+e^-\rightarrow Z^0Z^0)$ \cite{WW}.

\subsection{New particle searches}
The SM has some intrinsic inconsistencies and too many free parameters.
It could thus be an effective theory valid in the presently explored
energy scale.
Very many searches for new physics beyond the SM have been performed, 
without any positive indication 
\cite{particle-searches-at-lep},\cite{Igo-Kemenes},\cite{susy},\cite{susy2}.
We shall briefly summarize some searches.\\
{\bf Searches for SUSY particles.}
In supersymmetric (SUSY) models, each normal particle has a supersymmetric
partner whose spin differs by half a unit.
A new multiplicative quantum number, $R$-parity = $(-1)^{2S+3B-L}$,
is introduced, with value +1 for the SM particles and -1 for the SUSY partners.
If $R$-parity is conserved, SUSY particles are produced in pairs and they decay
to the lightest SUSY particle (LSP),
which may be the lowest
mass neutralino $\tilde{\chi}^0_1$.
In the Minimal Supersymmetric extension of the Standard Model (MSSM) all
sparticle masses are determined by 5 parameters:
$m_0$ = common mass at the GUT scale,
$M_2$ = SU(2) gaugino mass at the EW scale,
$\mu$ = mixing parameter of the two Higgs doublets,
$A$ = SUSY trilinear coupling.
Experimental limits depend on the choise of these 
parameters.
\par\noindent
\underline{Higgs bosons.}
In the MSSM there are 5 scalar Higgs bosons, $h^0$, $H^0$, $A^0$, $H^+$,
$H^-$;
the neutral ones are searched for
with methods similar to those used for the SM
Higgs boson; the limits are at the level of those for $H^0_{SM}$.
\par\noindent
\underline{Charginos.}
The fermionic SUSY partners of the $W^\pm$ and of the charged Higgs bosons
$H^\pm$ mix to form two mass eigenstates for each charge sign,
the charginos $\tilde{\chi}_{1,2}^\pm$.
They could be pair produced through $\gamma$ or $Z^0$ exchange in the
$s$-channel, or through sneutrino exchange in the $t$-channel.
The decays yield a neutralino,
$\tilde{\chi}^\pm \rightarrow \tilde{\chi}_1^0$ + ....
The experimental signature for pair production and decay is:
(i) two acoplanar leptons,
(ii) one lepton and one jet,
(iii) multi--jets with missing energy/momentum
(carried by neutralinos).
The existing limits are essentially at the kinematical limit,
$m_{\tilde{\chi}^\pm} > 103.5$ GeV \cite{susy}.
\par\noindent
\underline{Charged sleptons.}
Each SM lepton has two scalar partners, the right and left--held sleptons,
$\tilde{l}_R$ and $\tilde{l}_L$.
They could be pair produced through $s$-channel $\gamma$ or $Z^0$ exchange
or through $t$-channel neutralino exchange.
The main charged slepton decay is
$\tilde{l}^\pm \rightarrow \tilde{l}^\pm +\tilde{\chi}^0_1$.
 Mass limits are: $ m_{\tilde e} >$ 99.4 GeV,
$ m_{\tilde \mu} >$ 96.4 GeV, $ m_{\tilde \tau} >$ 87.1
GeV \cite{susy}.
\par\noindent
\underline{Scalar quarks.}
A scalar quark, in case of no mixing,
could be the lightest charged SUSY particle.
The dominant decay mode would be
$\tilde{t}\rightarrow c+\tilde{\chi}^0_1$.
The event topology would be two acoplanar jets with missing energy/momentum.
The limits, assuming a large $\theta_{\mathrm{mix}}$, are
$m_{\tilde{t}}>$ 95 GeV.
\par\noindent
\underline{Neutralinos.}
The $\tilde\gamma$, $\tilde Z^0$, $\tilde h^0$,$\tilde H^0$
mix to form 4 mass eigenstates, the neutralinos,
$\tilde{\chi}^0_i$, i=1,2,3,4.
The neutralinos could be pair produced through $s$-channel $Z^0$ exchange or
$t$-channel scalar electron exchange.
Since the lowest mass
$\tilde{\chi}_1^0$ is experimentally unobservable at LEP, the only way to look
for $\tilde{\chi}^0_1 \tilde{\chi}^0_1$ production is via the reaction
$e^+e^-\rightarrow \tilde{\chi}^0_1 \tilde{\chi}^0_1 \gamma$
or via $\tilde{\chi}^0_2 \tilde{\chi}^0_1$ pair production with
$\tilde{\chi}^0_2 \rightarrow \tilde{\chi}^0_1 l^+l^-$.
The limits obtained are at the level of $m_{\tilde{\chi}^0_1} >$ 40 GeV.
Higher mass limits are valid only for specific values of the SUSY parameters.
\par\noindent
\underline{R-Parity violation.}
If $R$-parity is violated, sparticles could be produced either in pairs or
singly and there are no constraints on the nature and on the stability of the
LSP.
If the LSP decays promptly, the event final states would be characterized by a
large fermion multiplicity.
If the LSP has a sizeable lifetime one would be able to directly observe
sparticles crossing the detector.
Both types of searches have been performed and no positive indication
was reported \cite{susy2}. Limits are quoted in the context of specific models.\\
{\bf Heavy charged and neutral leptons.}
The interest in the search for heavy neutral leptons has increased in
view of the possibility that neutrinos have non--zero masses
\cite{astrophysics}.
The typical event topology for pair production and decay
($e^+e^-\rightarrow N_l\bar{N}_l\rightarrow lW\bar{l}W$)
of unstable heavy neutral leptons would be two isolated leptons and 
two jets of hadrons.
Searches for long--lived charged heavy leptons,
$e^+e^-\rightarrow L^+L^-$,
involve topologies with back--to--back charged tracks.
For these searches one uses the central tracking detectors and
$dE/dx$ measurements.
The searches for long--lived neutral leptons assume
~$e^+e^-\rightarrow L^+L^-$,~ $L^\pm \rightarrow L^0W^\pm$.
The signature is a pair of acoplanar particles and missing transverse momentum.
Present limits extend to masses of about 100 GeV.\\
{\bf Excited fermions. Compositness.}
Composite models predict the existence of excited fermions,
$F^*$.
They are assumed to have the same electroweak
SU(2) and U(1) gauge couplings to the
vector bosons ($g$, $g\prime$) as the SM fermions;
but they are expected to group into left and right--handed weak isodoublets.
Excited fermions could be produced in pairs,
$e^+e^-\rightarrow F^{+*}F^{-*}$ or singly 
$e^+e^-\rightarrow F^{+}F^{-*}$.
For photonic decays, $F^*\rightarrow f\gamma$, the final states involve two
leptons and two photons.
For the $\bar{\nu}^*\nu^*$ case,
the final state involves 2$\gamma$ plus missing energy/momentum.
The present mass limits for singly produced excited fermions 
extend to $\sim$ 202 GeV.\\
{\bf Leptoquarks.}
Leptoquarks (LQs) are predicted in models which try to explain formally 
the symmetry between quarks and leptons; they could be produced 
in pairs and each decays into lepton + quark. At LEP present 
mass limits are at the level of $> 100$ GeV.\\
\par
In conclusion:
no evidence has been found for particles beyond the SM.
The searches will continue at all high energy colliders.

\section{Lepton--nucleon collisions}
\begin{figure}[htb]
\begin{center}
 \mbox{
    \epsfig{figure=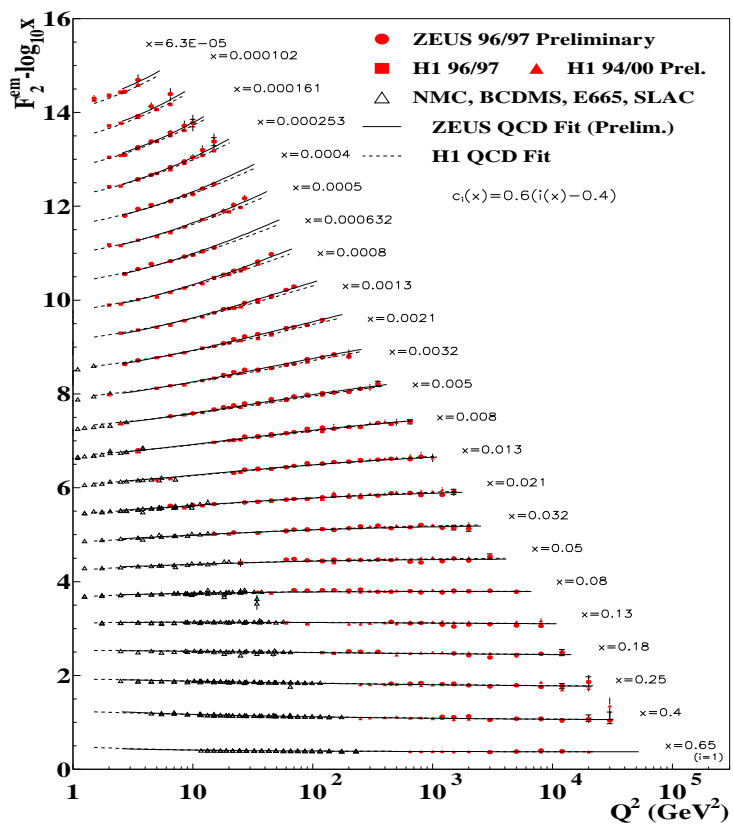,height=11truecm,
           bbllx=80,bblly=150,bburx=465,bbury=570}
 }
\end{center}
\caption{ $F_{2}(x,Q^{2})$ at fixed $x$ values as a function 
               of $Q^2$ from different experiments.}
\label{fig:DIS}
\end{figure}

The HERA Collider at DESY in Hamburg is an asymmetric $ep$ collider, 
$E_e \cong 30$ GeV, $E_p \cong 820$ GeV, $E_{cm} \cong 300$ GeV.  
Two general purpose detectors, H$_1$ and ZEUS, are taking data at 
luminosities $\cal L$ $\sim 10 ^{31}$ cm$^{-2}$ s$^{-1}$ \cite{hera}.
One of the main physics purposes of these experiments is the study of 
DIS, either via $\gamma$, $Z^0$ exchange 
($ep \rightarrow eX$) or via $W^\pm$, $ep \rightarrow \nu_e X$.\\ 
\indent
In the quark--parton model the ``current particle" ($\gamma$, $Z^0$, 
$W^\pm$) emitted by the incoming electron interacts with one of the 
quarks of the proton. The scattered quark gives rise to a jet of hadrons
(current jet). The proton remnants give rise to a second jet (target jet).
The process may be described in terms of the four momentum transferred squared 
$Q^2$ and the energy transferred $\nu = E_e - E_{e'}$ (or the variable 
$x = Q^2/ 2m_p \nu =$ fraction of the 
proton momentum carried by the struck quark, and $y = Q^2/ xs$).\\
\indent
The cross--section for NC deep inelastic $ep$ scattering may be computed 
from the elastic electron--quark cross--section. In the quark--parton 
model the only unknown is the probability $q(x)$ for finding a quark 
$q$ in the proton carrying a fraction $x$ of the proton momentum.
The structure function $F^p_2$ is $F^p_2 = \Sigma ~e^2_q 
~ q(x)$. $F_2$ should be independent of $Q^2$ 
(Bjorken scaling). For NC deep inelastic scattering the differential 
cross--sections depend on $F_2(x,Q^2)$. \\
\indent
Because of the large c.m. energy, HERA allows to study the
structure function $F_2(x,Q^2)$ at very small values  of x and at very large 
$Q^2$ \cite{hera}. Fig. \ref{fig:DIS} shows a recent compilation of $F_2$ plotted 
versus $Q^2$ for different values of $x$. 
The variation of $F_2$ with
$Q^2$ (scale breaking) is predicted by QCD from gluon bremsstrahlung 
and quark pair creation by gluons. The data are in agreement with this 
picture. $F_2$ increase considerably at small values of $x$ for any 
value of $Q^2$. 
The data indicate that one half of the proton momentum is carried by
gluons, which dominate for $x < 0.2$. Even if gluons contribute 
$50\%$ of the proton momentum, it is difficult to extract the gluon 
density function $g(x,Q^2)$ because gluons do not contribute 
directly to DIS in the quark--parton model.   
One can extract the gluon structure function from the slope of 
$dF_2 / d$ln$Q^2$, obtaining $g(x) \sim x^{-(1+\lambda)}$
with $\lambda \sim 0.5$. Thus the number of partons in the proton 
increases at low values of $x$ (and also at high $Q^2$ 
for a fixed x value $<0.1$).
This means that when we look inside the proton with a better 
resolution we see more partons, and the 
number of gluons tends towards a large value.
HERA allows studies of photoproduction and of diffractive processes.
It also allows to search for many new particles; the limits are particularly 
interesting for leptoquarks: $m_{LQ} > 290$ GeV for first generation scalar 
leptoquarks \cite{herasearch}.

\section{Hadron--hadron collisions}
Most of the recent experimental results on high energy 
hadron--hadron ($hh$) collisions come from fixed target 
experiments at CERN, Fermilab and Serpukhov. The upgraded Fermilab
$p\bar p$ collider started recently data taking at $E_{cm} = 2$ TeV 
with luminosities $\cal L$ $>10^{31}$ cm$^{-2}$ s$^{-1}$.
Two general purpose upgraded detectors, CDF and D0, are taking data
mainly on large $p_t$ physics.
\begin{figure}[thb]
\begin{center}
 \mbox{
    \epsfig{figure=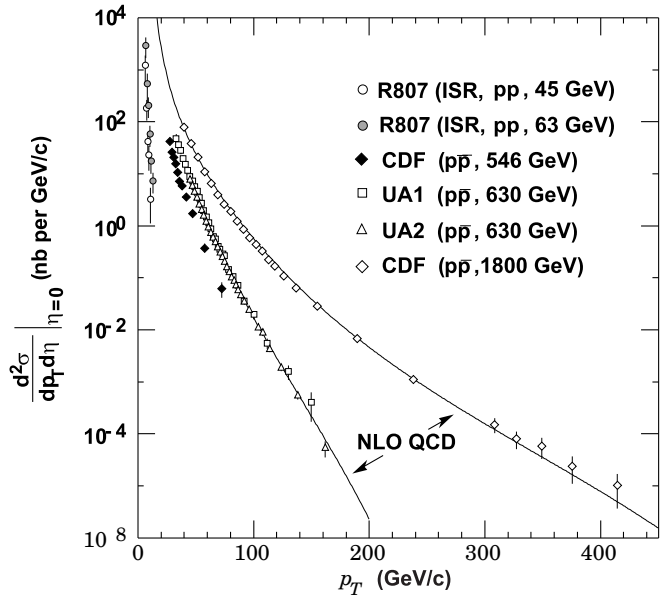,height=8.0truecm,
            bbllx=250,bblly=300,bburx=510,bbury=525}
 }
\end{center}
\caption{Differential inclusive jet cross--sections at $\eta =0$ 
         in $hh$ collisions at different c.m. energies.}
\label{fig:CDF}
\end{figure} 
\par\noindent
{\bf Elastic and total cross--sections. Low $p_t$ physics.}
The total $p\bar p$ cross--section may be written as:
$\sigma_{tot} = \sigma_{el} + \sigma_{inel} = \sigma_{el} + 
\sigma_{sd} + \bar\sigma_{sd} + \sigma_{dd} + \sigma_{nd}$,
where $\sigma_{el}$ is the elastic cross--section, $\sigma_{sd}$ 
is the single diffractive
cross--section when the incoming proton fragments into a number of 
particles, $\bar\sigma_{sd}$ is the single diffractive cross--section 
for the fragmentation of the antiproton (at high energies $\sigma_{sd} = 
\bar\sigma_{sd}$), $\sigma_{dd}$ is the double diffractive cross--section,
$\sigma_{nd}$ is the non--diffractive part of the inelastic cross--section.
The elastic, single diffractive and double diffractive processes
give rise to low multiplicity events with particles emitted in the very
forward and very backward regions in the c.m. system.
The non--diffractive cross--section is the main part of the inelastic 
cross--section; non--diffractive processes give rise to high  multiplicity 
events and to particles emitted at all angles. Most of the non--diffractive 
cross--section concerns particles emitted with low transverse momentum
($low$ $p_t$ $physics$) with properties which change slowly with c.m. energy 
($ln s$ $physics$).
\par
The total cross--sections for $p\bar p$, $pp$, $\pi^-p$, $\pi^+p$, 
$K^-p$, $K^+p$ 
collisions for c.m. energies between 3 and 70 GeV indicate that as the 
energy increases, the cross--sections decrease, reach a minimum and then
increase. The $K^+p$ cross--section increases in this entire energy range:
its rise was already evident and established at Serpukhov energies.
All the differences $\sigma_{tot}(\bar xp) - \sigma_{tot}(xp)$ 
decrease with increasing energy\cite{total-xsection}.
The slope parameter $b$ for $p\bar p$ elastic scattering keeps 
increasing up to $\sqrt{s} = 1.8$ TeV, at a rate which is probably
larger than ln$s$. The total elastic $p\bar p$ cross--section 
continues to rise at a rate close to ln$^2 s$ or larger. Therefore
the ratio $\sigma_{el}/\sigma_{tot}$ increases with energy,
reaching the value of about 0.24 at $\sqrt{s} = 1.8$ TeV.
This value is still below the black disk value of 0.5, though the 
center of the proton has become more opaque, almost black. 
Present values of $\rho$, the ratio of the real to the imaginary 
part of the $p\bar p$ elastic scattering amplitude, obtained from 
elastic scattering data in the Coulomb--Nuclear interference region, 
follow the dispersion relation predictions.\\
\indent
The average charged particle multiplicity has increased with energy
reaching a value $<n_{ch}> \cong  40$ at $E_{cm} \cong 1.8$ TeV.\\  
\indent 
In most elastic scattering analyses the polarization effects are neglected:
such effects are thought to decrease as the energy increases 
and they are small in the forward direction. But sizable and 
measurable spin and asymmetry effects were observed at 
$p_{lab} = 28$ GeV/c and at high $p_t$, where the
cross section for $pp$  elastic scattering with spins parallel is 
about four times larger than the cross section for spins antiparallel
 ($p_t \sim 2 - 3$ GeV/c). Spin effects need to be investigated 
more systematically, both experimentally and theoretically.\\
{\bf Large $p_t$ physics.}
 A relatively small part of the non--diffractive cross--section
is due to central collisions among the colliding particles and gives rise to 
high $p_t$ jets of particles emitted at large angles ($large$ $p_t$ $physics$).
The contribution of jet physics increases with c.m. energy. 
The two--jet production in $p\bar p$ collisions, $p+\bar p \rightarrow 
2$ jets $+ X$, is due to parton--parton processes \cite{SFM}. Fig. \ref{fig:CDF}
shows a compilation of inclusive jet cross--sections measured 
in hadron--hadron collisions at different c.m. energies.
Good agreement is found with the QCD prediction, thus excluding new 
physics, such as that expected from subconstituent contact interactions \cite{CDF}.\\
{\bf The top quark.}
The CDF and D0 collaborations found the top quark and are studying the detail of 
its production and decay \cite{top}. In $p\bar p$ collisions the dominant 
channels are via quark--antiquark annihilation 
or gluon--gluon fusion: $q + \bar q \rightarrow t + \bar t$, ~~~$g + g 
\rightarrow t + \bar t$. The dominant decays of the top quark are 
\begin{center}
$t \rightarrow W^+ + b$,~~~ $t \rightarrow W^+ +$ ($s$ or $d$),~~~
$t \rightarrow g + W^+ +$ ($b$, $s$ or $d$) 
\end{center}
 and likewise for $\bar t$. The present $t$ mass value is quoted 
in Fig. \ref{fig:average}.\\
{\bf New particle searches.}
The CDF and D0 experiments give 95\% CL limits on many new particles,
in particular: $m_{LQ} > 160 (148)$ GeV for second (third) generation scalar LQs,
$m_{Z'} > 670$ GeV, $m_{W'} > 755$ GeV \cite{cdfsearch}.

\section{Conclusions}
Experiments at the large colliders (LEP, SLC, Fermilab, HERA)
and at lower energies collected an impressive amount 
of data, which together with new theoretical calculations, provide
stringent tests of the Standard Model.
\par
The main physics results may be summarized as follows: three
neutrino families, lepton universality, precise determination of 
electroweak and strong parameters ($m_Z$, $\Gamma_Z$, sin$^2 \theta_W$,
$m_W$, $m_t$, $\alpha_s$, .....), the flavour independence of $\alpha_s$,
the running of the strong and of the E.M. coupling costants, the
existence of the triple  boson vertex $Z^0W^+W^-$, the determination 
below threshold of the mass of the top quark and of its discovery above
threshold, possible indications and limits on the $H^0_{SM}$ mass, 
$b$ and $\tau$ physics, precise measurements of the lifetimes of short 
lived particles, studies of QCD at large and small $p_t$, detailed studies 
of multihadron final states and of the difference between quark and gluon jets,
the deeper structure of the proton and of the photon, etc.
All experiments gave increasingly better limits on new particles 
and new phenomena.
Let us hope that the new increases in energy and luminosity will
really open up a new field.
These results have also strong implications in the astroparticle 
physics field \cite{astrophysics}.
\par
Experiments at lower energies provided information on direct CP violation
in the $K^0\bar{K}^0$ system and of CP violation in $B^0\bar{B}^0$,
have shown the importance of understanding the spin of the proton 
(spin crisis) and of polarization measurements, the
first indications for the quark--gluon plasma, etc.
\par
One should not neglect the very large number of Diploma, Laurea
and PhD theses using high energy data and the strong impact 
of fundamental high energy physics on the public understanding of science.\\
\par
We would like to thank the members of the OPAL Collaboration for their
cooperation;
we acknowledge the collaboration of D. Bonacorsi, F. Fabbri, P. Giacomelli, 
S. Marcellini, 
and other members of the
Bologna groups.

\end{document}